\documentclass[aps, prb, fleqn, superscriptaddress]{revtex4} % Physical Review B

\usepackage[colorlinks = true,
linkcolor = blue,
urlcolor  = blue,
citecolor = blue,
anchorcolor = blue]{hyperref}

\usepackage{ragged2e}

\newcommand{\bt}{BaTiO$_3$}

\newcommand{\fo}{Fe$_3$O$_4$}

\usepackage{float}
\usepackage{graphicx}
\usepackage[dvips]{epsfig}
\usepackage{bm, amssymb}
\usepackage{dcolumn}
\usepackage{xcolor}

\begin{document}

\title{Strain and electric field control of magnetic and electrical transport properties in a magneto-elastically coupled \fo/\bt (001) heterostructure}

\author{Gyanendra Panchal$^{*}$}
\affiliation{Department Dynamics and Transport in Quantum Materials, Helmholtz-Zentrum Berlin f\"{u}r Materialien und Energie, Hahn-Meitner-Platz 1, 14109 Berlin, Germany}
\affiliation{UGC DAE Consortium for Scientific Research, University Campus, Khandwa Road, Indore 452001, India}
\email[corresponding authors (Gyanendra Panchal and Ram Janay Choudhary): ]{gyanendra.panchal@helmholtz-berlin.de, ram@csr.res.in}

\author{Danny Kojda}
\affiliation{Department Dynamics and Transport in Quantum Materials, Helmholtz-Zentrum Berlin f\"{u}r Materialien und Energie, Hahn-Meitner-Platz 1, 14109 Berlin, Germany}

\author{Sophia Sahoo}
\affiliation{UGC DAE Consortium for Scientific Research, University Campus, Khandwa Road, Indore 452001, India}
\author{Anita Bagri}
\affiliation{UGC DAE Consortium for Scientific Research, University Campus, Khandwa Road, Indore 452001, India}
\author{Hemant Singh Kunwar}
\affiliation{UGC DAE Consortium for Scientific Research, University Campus, Khandwa Road, Indore 452001, India}

\author{Lars Bocklage}
\affiliation{Deutsches Elektronen-Synchrotron DESY, Notkestraße 85, 22607 Hamburg, Germany}

\affiliation{The Hamburg Centre for Ultrafast Imaging, Luruper Chaussee 149, 22761 Hamburg, Germany}

\author{Anjali Panchwanee}
\affiliation{Deutsches Elektronen-Synchrotron DESY, Notkestraße 85, 22607 Hamburg, Germany}

\author{Vasant G. Sathe}
\affiliation{UGC DAE Consortium for Scientific Research, University Campus, Khandwa Road, Indore 452001, India}

\author{Katharina Fritsch}
\affiliation{Department Dynamics and Transport in Quantum Materials, Helmholtz-Zentrum Berlin f\"{u}r Materialien und Energie, Hahn-Meitner-Platz 1, 14109 Berlin, Germany}

\author{Klaus Habicht}
\affiliation{Department Dynamics and Transport in Quantum Materials, Helmholtz-Zentrum Berlin f\"{u}r Materialien und Energie, Hahn-Meitner-Platz 1, 14109 Berlin, Germany}
\affiliation{Institute of Physics and Astronomy, University of Potsdam, Karl-Liebknecht-Str. 24-25, D-14476 Potsdam, Germany}

\author{Ram Janay Choudhary$^{*}$}
\affiliation{UGC DAE Consortium for Scientific Research, University Campus, Khandwa Road, Indore 452001, India}

\author{Deodutta M. Phase}
\affiliation{UGC DAE Consortium for Scientific Research, University Campus, Khandwa Road, Indore 452001, India}

%\date{\today}

\begin{abstract}

We present a study of the control of electric field induced strain on the magnetic and electrical transport properties in a magneto-elastically coupled artificial multiferroic Fe$_3$O$_4$/BaTiO$_3$ heterostructure. In this Fe$_3$O$_4$/BaTiO$_3$ heterostructure, the Fe$_3$O$_4$ thin film is epitaxially grown in the form of bilateral domains, analogous to \textit{a-c} stripe domains of the underlying BaTiO$_3$(001) substrate. By \textit{in-situ} electric field dependent magnetization measurements, we demonstrate the extrinsic control of the magnetic anisotropy and the characteristic Verwey metal-insulator transition of the epitaxial Fe$_3$O$_4$ thin film in a wide temperature range between 20-300 K, via strain mediated converse magnetoelectric coupling. In addition, we observe strain induced modulations in the magnetic and electrical transport properties of the Fe$_3$O$_4$  thin film across the thermally driven intrinsic ferroelectric and structural phase transitions of the BaTiO$_3$ substrate. \textit{In-situ} electric field dependent Raman measurements reveal that the electric field does not significantly modify the anti-phase boundary defects in the Fe$_3$O$_4$ thin film once it is thermodynamically stable after deposition and that the modification of the magnetic properties is mainly caused by strain induced lattice distortions and magnetic anisotropy. These results provide a framework to realize electrical control of the magnetization in a classical highly correlated transition metal oxide.
\end{abstract}

\maketitle

\section{Introduction}
Magnetoelectric coupling effects in artificial multiferroic heterostructures have attracted huge attention over the last few years, because of strong coupling between ferroelectricity and magnetism and its potential for next generation spintronics and memory device applications.\cite{Bader2010,Martin0012,Scott2007} Recently, electric field control of the magnetization, which is referred to as converse magnetoelectric (ME) effect, has become increasingly important in spintronics, where it is essential for device operation with very low energy consumption.\cite{Fusil2014,Vaz2012,Taniyama2015} 
It has been shown that artificial multiferroic horizontal heterostructures made of ferromagnetic (FM) and ferroelectric (FE) materials are most promising for the realization of electric field control of the magnetization via a strain-mediated converse ME effect.\cite{Ramesh2007a,Eerenstein2006} In such magnetoelastically coupled FM/FE heterostructures, an electric field induces strain in the FE material (via converse piezoelectric effect), which is transferred into the FM layer where it modifies the magnetization via converse magnetoelastic effects.\cite{YWANG}

In this work, we present a study of the influence of strain and external electric field on the magnetic and electrical transport properties in a magneto-elastically coupled heterostructure built of a magnetite, \fo\ thin film grown on a \bt\ (BTO) substrate.

At room temperature, BTO is a well known lead free FE perovskite oxide with tetragonal (T) crystal structure (space group \textit{P}4\textit{mm}, \textit{a} = \textit{b} = 3.99 \AA\ and \textit{c} = 4.03 \AA) and a Curie temperature $T_{\text{C}}$ = 393 K. During cooling, it undergoes two structural (and also ferroelectric) phase transitions to orthorhombic (O) and rhombohedral (R) phases at 278 K and 183 K, respectively. Therefore, in FM/FE heterostructures based on a BTO substrate, strain at the FM/FE interface can be tuned either by the intrinsic BTO structural phase transitions or by extrinsic electric field induced lattice distortions. These two additional degrees of freedom along with strong elastic coupling at the uniform FM/FE interface makes BTO a suitable FE material to design artificial planar FM/BTO(FE) heterostructures. As such, many efforts in the last decades have been devoted to understand the converse ME effect\cite{Eerenstein2007,Motti2018} and to control the strain-induced modifications of the magnetic and electrical transport properties in FM/BaTiO$_3$ heterostructures, where different FM layers such as FePt,\cite{Weisheit2007} CoFe,\cite{Lahtinen2013} La$_{0.67}$Sr$_{0.33}$MnO$_3$,\cite{Lee2000} Sr$_2$CrReO$_6$,\cite{Czeschka2016} FeRh,\cite{Bennett2016} or La$_{0.67}$Ca$_{0.33}$MnO$_3$\cite{Moya2012} have been investigated.

As FM component, magnetite or Fe$_3$O$_4$, has also been used recently in FM/BTO heterostructures.\cite{Lu2019,Vaz2009,tian2008} It is a most promising quantum material for device applications, because of its outstanding properties such as room temperature ferrimagnetic nature, high spin polarization along with half metallicity,\cite{Dedkov2002,Fonin2007} and characteristic Verwey metal-insulator transition at around 120 K.\cite{Verwey1939, Liu2016} Despite probably being the most studied magnetic material in the last century, the mechanisms of strain mediated converse ME coupling\cite{Vaz2009,tian2008} and of how an applied electric field influences the Verwey metal-insulator transition, the mechanisms of charge ordering below the Verwey transition and the magnetic anisotropy in Fe$_3$O$_4$ are still not fully understood and remain an active area of research. Recently, Wong \textit{et al.}\cite{Wong2012} demonstrated the manipulation of the Verwey transition of Fe$_3$O$_4$ thin films by electrical gating. Zhang \textit{et al.}\cite{Zhang2018}, shown that the Verwey transition temperature of an epitaxial Fe$_3$O$_4$ film in liquid/Fe$_3$O$_4$/MgO heterostructure is strongly dependent on the ionic liquid gate voltage.

In this work, we have grown an epitaxial horizontal multiferroic Fe$_3$O$_4$/BTO(001) heterostructure and show electrical control of the Verwey transition and magnetic anisotropy via $in$-$situ$ application of an electric field over a broad temperature range from 20 K to room temperature under strong converse ME coupling. By performing temperature dependent magnetization and resistivity measurements, we also demonstrate the strain driven manipulation of magnetization and electrical transport properties of the elastically coupled Fe$_3$O$_4$ thin film across the structural phase transitions of the underlying ferroelectric BTO substrate. Room temperature electric field dependent Raman measurements are also carried out to study the effect of electric field on the presence of anti-phase boundaries (APBs) and related magnetic properties.

\section{Experimental details}
An epitaxial thin film of \fo\ was grown on a single crystalline BTO(001) substrate by Reflection High-Energy Electron Diffraction (RHEED) assisted pulsed laser deposition using a KrF excimer laser (248 nm). A single-phase stoichiometric $\alpha$-Fe$_2$O$_3$ target was used for the deposition, which was carried out in vacuum (1x10$^{-6}$ Torr) at a substrate temperature of 445 $^\circ$C with 6 Hz repetition rate. The growth has been monitored \textit{in-situ} by RHEED oscillations (STAIB Instruments). During the deposition, the energy density of the laser beam at the target surface was kept around 2.0 J/cm$^{2}$. 
For the structural characterization of the thin film, Reciprocal space mapping (RSM) at room temperature was carried out using a high-resolution Bruker D8-Discover X-ray diffractometer. DC magnetization measurements were performed in a 7-Tesla SQUID-vibrating sample magnetometer (SVSM; Quantum Design Inc., USA) in the temperature range between 20 – 320 K. Room temperature magneto-optical Kerr effect (MOKE) microscopy was performed by Evico Magnetics Optical Kerr microscope. \textit{In-situ} electric-field dependent magnetic properties were studied by applying an electric field ($E$-field) up to 10 kV/cm across the BTO substrate along [001] direction using a custom built insert for the SVSM. Resistivity measurements as a function of temperature were carried out by measuring I-V curves using the van der Pauw method.\citep{vdp} During the measurements the cooling and heating rates were 0.1K/min. The contacts on the small piece of sample were formed by gold-aluminum wire via wire bonder 5630 (F$\&$K Delvotec Semiconductor GmbH) and excitation currents from +24 $\mu$A to -24 $\mu$A were used for the I-V curves. Room temperature Raman spectra as function of different electric fields were recorded with a He-Ne laser (632.8 nm) in back-scattering geometry using a homemade stage in a Horiba LABRAM spectrometer.

\section{Results and discussion}
\subsection{Structural properties}

\textit{In-situ} RHEED has been employed to monitor the surface and growth mode of the Fe$_3$O$_4$ thin film during deposition. Fig.1 (a) shows the RHEED oscillations (shown by arrows). These oscillations along with vertical stripes in the RHEED diffraction pattern (top inset of Fig. 1(a)) illustrate the layer-by-layer growth of the Fe$_3$O$_4$ thin film.\cite{Moyer2015} Based on these RHEED oscillations, the nominal thickness of the thin film is 38 nm.
To characterize the structural properties of the film, high-resolution reciprocal space mapping (RSM) was employed. Fig.1(b) shows the room temperature RSM of the Fe$_3$O$_4$/BTO heterostructure in the tetragonal phase of BTO around the symmetric (002) plane. In this map, the BTO substrate exhibits Bragg spots at different q$_z$ values, indicating the presence of 90$^\circ$ ferroelectric $a$-$c$ stripe domains as been previously observed.\cite{Lahtinen2011} At room temperature, these domains are usually formed at the surface of the BTO single crystal in the form of a ferroelectric stripe pattern which is clearly visible in the polarization microscopy image as shown in the bottom inset of Fig.1(a). The fractions of these domains are influenced by the thermal and electrical poling history of the BTO sample after thermal cycling across its cubic to tetragonal structural phase transition.\cite{Lahtinen2011} In the RSM measurement, the sample is aligned at a reflection from the $a$-domains of the BTO substrate, which results in an intense peak at q$_z$ = 4.962 nm$^{-1}$ from these $a$-domains and a weaker peak at q$_z$ = 4.906 nm$^{-1}$ from the $c$-domains. The calculated misorientation between these domains is around 0.6 degrees.

\begin{figure*}[t]
\centering
\includegraphics[width=0.55\textwidth]{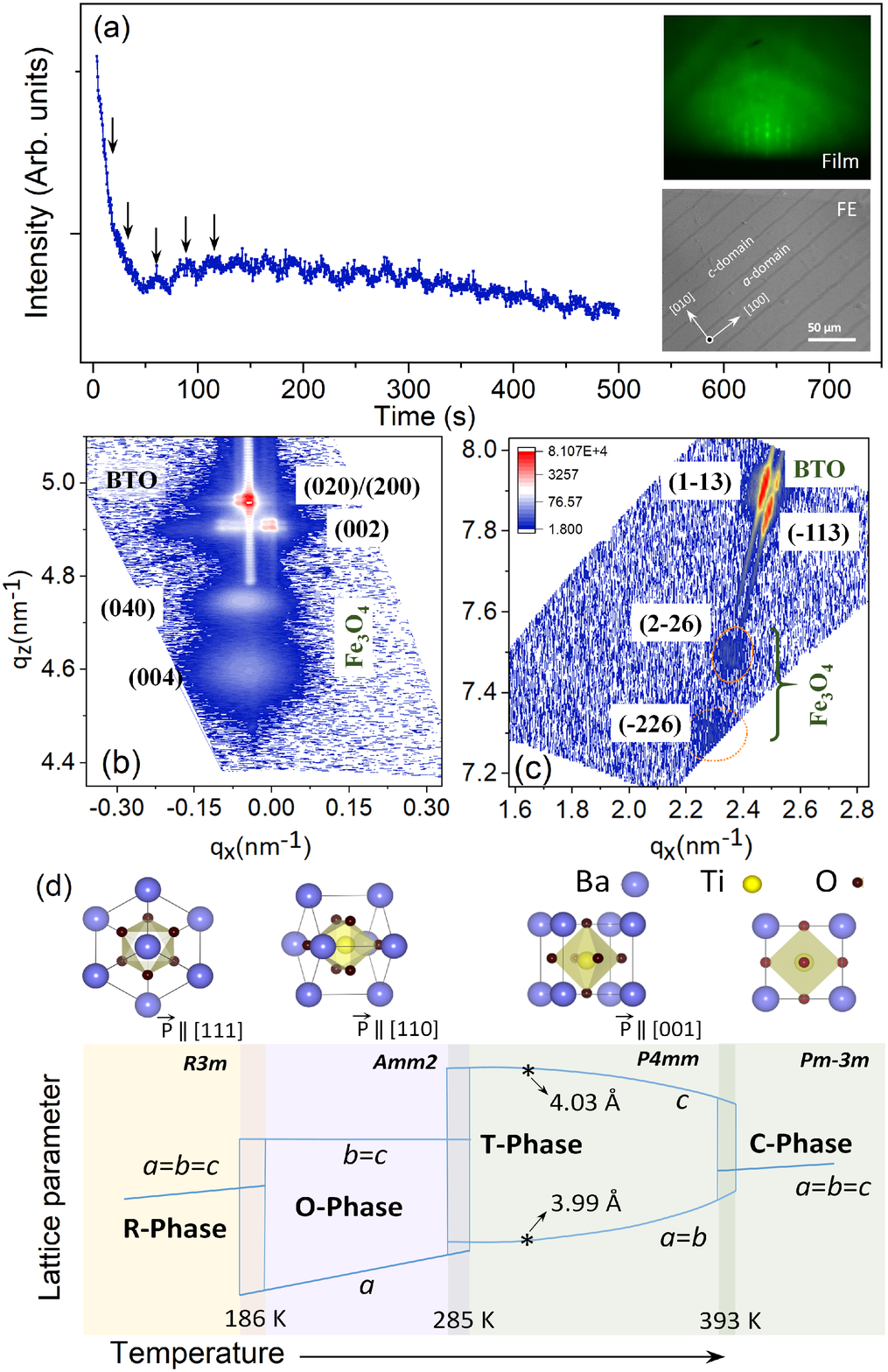}

\caption{(a) RHEED oscillations of Fe$_3$O$_4$ thin film growth on BTO(001) substrate. The top inset shows the RHEED pattern of the 38 nm Fe$_3$O$_4$ thin film grown on BTO(001) and the bottom inset shows the room temperature polarization microscopy image of 90$^\circ$ ferroelectric $a$-$c$ stripe domains. Reciprocal space mapping of the  Fe$_3$O$_4$/BTO heterostructure around the (b) symmetric (002) and (c) asymmetric (-113) planes of BTO. (d) Schematic representation of the structural evolution of BTO with temperature: Lattice parameters, crystal structure and direction of the ferroelectric polarization.\cite{kay}}   
\label{abc}
\end{figure*}

The presence of \textit{two} Bragg peaks [shown in Fig.1(b) and Fig.1(c)] which correspond to the Fe$_3$O$_4$ film in each of the scans, symmetric (002) and asymmetric (-113), reveals that the Fe$_3$O$_4$ film is epitaxially grown in the form of bilateral domains, analogous to the underlying BTO \textit{a-c} stripes domain pattern.\cite{Panchal2018} The thin film is partially relaxed \textit{in-plane} because of a high degree of lattice mismatch between Fe$_3$O$_4$ (cubic structure with \textit{a} = 8.396 $\AA$) and BTO (4.47 $\%$ with $c$-domain and 6.97 $\%$  with $a$-domain) lattice parameters.\cite{Phase2019} These results from X-ray diffraction measurements clearly confirm that the Fe$_3$O$_4$ thin film adopts the underlying domain pattern of the BTO substrate. 

\subsection{Magnetic properties}

The magnetization versus temperature M(T) behavior of the Fe$_3$O$_4$/BTO heterostructure has been studied in the temperature range 20-320 K under two non-saturating magnetic fields $B$ of 15 mT and 30 mT, which were applied along the in-plane direction [100] of the film. The data have been recorded during warming in zero field cooling conditions (ZFC), followed by cooling in field (FCC) and are shown in Fig.2(a). The magnetization exhibits several interesting features. Most notably, it exhibits large discontinuous jumps at the structural phase transitions of the BTO substrate. The associated abrupt variation of the lattice parameters and of the direction of ferroelectric polarization during the structural phase transitions of BTO is schematically represented in Fig. 1(d).\cite{kay} For $B$ = 15 mT, in the FCC cycle, the magnetization increases by $\sim$ 20 \% [{(M$_f$- M$_i$)×100/M$_i$}] when the temperature is decreased through  the T to O-phase transition. On further decreasing the temperature to the O-R phase transition, another magnetization rise of $\sim$ 17 \% is observed. These distinct magnetization jumps are also observed in the ZFC cycle, but with opposite sign and with larger magnitudes. Such different magnetization behavior of the FCC M(T) cycle (with respect to the ZFC cycle) is mainly a consequence of the field cooling (FC) process, which establishes a magnetic anisotropy along the field cooling direction during the field cooling and influences the magnetization jumps with respect to the ZFC M(T) cycle. The height and sign of the jumps strongly depend on the direction, strength of external cooling field and the strength of applied magnetic field for data recording. The presence of sharp magnetization jumps in the M(T) behavior at the transition temperatures of BTO demonstrates that the magnetization is modulated by the strain induced lattice distortion at the interface caused by the structural phase transition of BTO.

\begin{figure*}[t]
\includegraphics[width=0.60\textwidth]{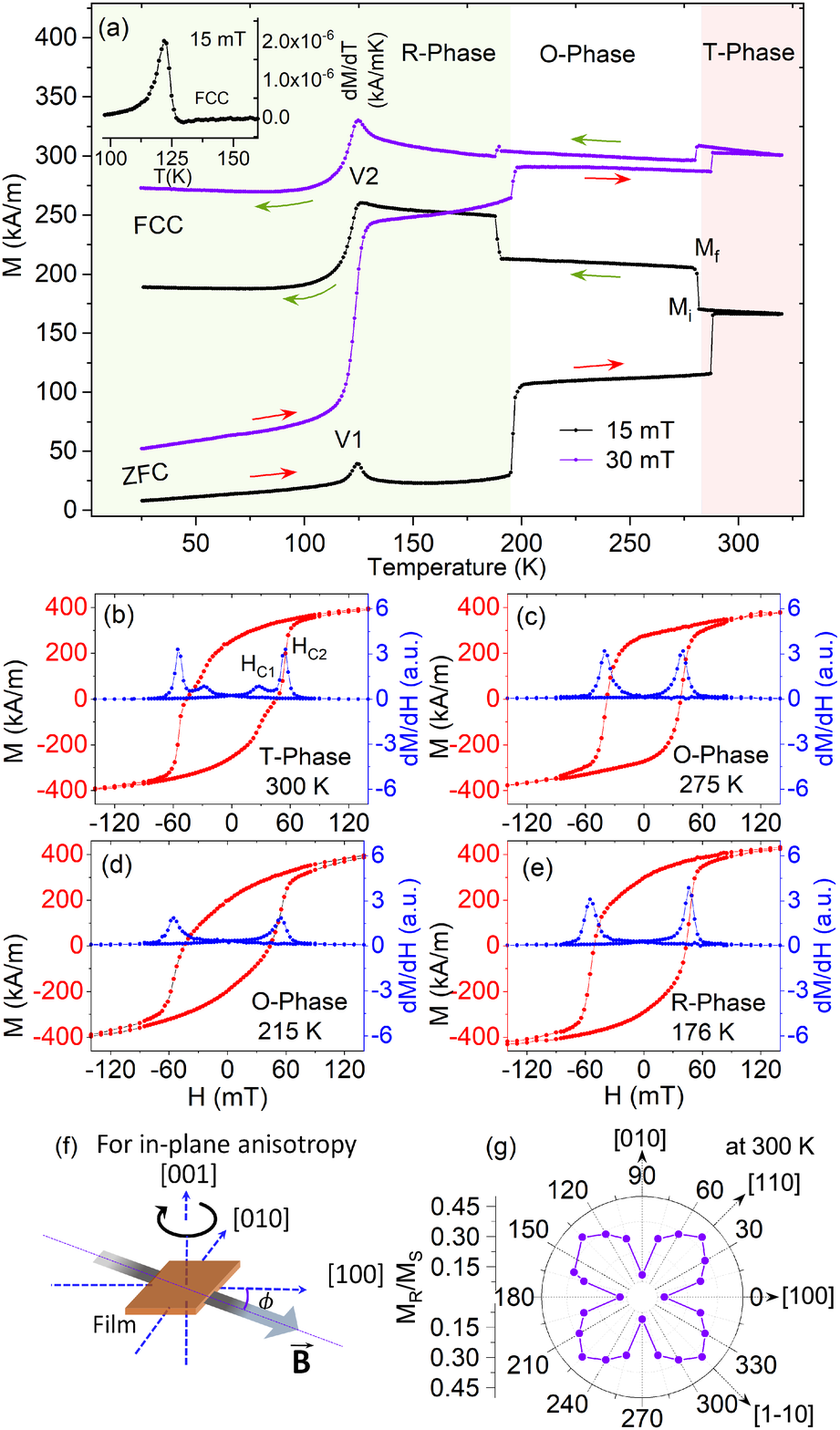}

\caption{(a) M(T) behavior under $B$= 15 and 30 mT field cooled cooling (FCC) and zero field cooled (ZFC) cycle. The inset shows the derivative $dM/dT$ across the Verwey transition. [(b)-(e)] Magnetization hysteresis M(H) loop and derivative $dM/dH$ of Fe$_3$O$_4$/BTO(001) heterostructure for different temperatures corresponding to the different structural phases of the BTO substrate. (f) Schematic for in-plane angle dependent MOKE measurements (g) Polar plot of normalized remanent magnetization (M$_R$/M$_S$) at room temperature (T-phase).}
\label{fig2}
\end{figure*}

Along with these jumps, the characteristic Verwey transition of \fo\ is clearly visible at $T$ = 123 K [inset of Fig.2(a)], which confirms the growth of a single-phase stoichiometric Fe$_3$O$_4$ thin film.
For a larger applied field of $B$ = 30 mT, the profile of the Verwey metal-insulator transition gets modified [V1 and V2 in Fig.2(a)]. The magnetization jumps across the structural transitions in ZFC conditions follow the behavior in lower magnetic field, but their magnitudes are suppressed. Interestingly, under FCC conditions, the signs of the magnetization jumps are inverted with respect to those at $B$ = 15 mT, as shown in the in Fig.2(a). This behavior can be explained by considering the competition between magnetic and elastic energy by the lattice distortion induced strain. As a larger magnetic field is applied to the system, the competing magnetic energy starts to dominate over the elastic energy, which results in the suppression of the magnetization jumps across the structural transitions. The inversion of the sign of the jumps in larger magnetic field and the variation of the magnitude of these distinct jumps around the structural phase transitions of BTO suggest that the strain induced lattice distortion also alters the magnetic anisotropy and changes the direction of the magnetic easy axis. Similar observations have previously been reported for the La$_{0.7}$Sr$_{0.3}$MnO$_3$/BTO(001) system.\cite{Panchal2018JM}

To understand the changes in the magnetic anisotropy systematically, we have recorded magnetization hysteresis M(H) loops of the Fe$_3$O$_4$/BTO heterostructure around the structural transitions of the BTO substrate. Each M(H) loop was recorded after zero magnetic field cooling from 320 K to the desired temperatures (300, 275, 215, and 176K) corresponding to the different structural phases (tetragonal, orthorhombic and rhombohedral phases) of the underneath BTO substrate. Different strain state dependent M(H) hysteresis curves of the Fe$_3$O$_4$ thin film in the tetragonal (T-phase at 300 K), orthorhombic (O-phase at 275-215 K) and rhombohedral (R-phase at 176 K) structural phases of BTO are shown in Figs.2(b)-(e). In the T-phase of the BTO, the M(H) exhibits a double coercivity behavior related to hard and soft magnetic-like phases (H$_{C1}$ for soft and H$_{C2}$ for hard), clearly visible in the derivative $dM/dH$ as shown in Fig.2(b). This behavior can be explained as follows:  The Fe$_3$O$_4$ thin film is grown on an imprinted 90$^\circ$ \textit{a-c} domain pattern from the bottom BTO substrate [Fig.2(b)]. This room temperature multidomain state then creates two strain regions of different magnetic anisotropy energy in the Fe$_3$O$_4$ thin film and gives rise to a double coercivity hysteresis M(H) in T-phase. This observation is fully consistent with recent reports of numerical simulations on the ME-effect in an epitaxially grown Fe$_3$O$_4$/BTO heterostructure, based on a two-region model consisting of two magnetic regions related to the underlying ferroelastic \textit{a} and \textit{c} domain state of BTO.\cite{Geprags2013}

Upon entering into the orthorhombic (O) phase with decreasing temperature, the magnetic anisotropy is modified and the double coercivity-like behavior completely vanishes with a large reduction 19 \% in coercivity from 46.3 mT (at 300 K) to 37.3 mT (at 275 K) as shown in Fig.2(c). This suggests that our BTO substrate changes from its higher-temperature multidomain state to a rectangular domain state\cite{Vaz2009} when it is passing through the T-O structural phase transition. We also observe a 13 \% increment in the remanence ratio M$_R$/M$_S$ and a small variation in the saturation magnetization in that phase, which implies that the magnetization reversal becomes more coherent. These observations are closely related to the modification of the strain state and of the density of the anti-phase boundaries (APBs) in Fe$_3$O$_4$ thin films. In fact, recently X. H. Liu \textit{et al.}\cite{Liu2017} investigated the influence of different substrate induced lattice mismatches on the magnetic properties of Fe$_3$O$_4$ thin films and have shown that the density of APBs and the demagnetization process are strongly affected by the lattice mismatch induced strain states. 

During further cooling in the O-phase of BTO, the in-plane lattice parameter \textit{`a'} of $c$-domains (majority domain fraction) starts decreasing\cite{Lee2000} and increases the strain or lattice mismatch between Fe$_3$O$_4$ and BTO. This leads to an incoherent or less coherent magnetization reversal as observed in Fig.2(d). When the temperature is further decreased into the R-phase, the lattice parameter \textit{a} then rapidly increases and reduces the transferred strain across the transition (see Fig.1(d)). This further modifies the magnetic anisotropy as evidenced by the broadened M(H) hysteresis curve with an 18 \% increased coercivity with respect to the O-phase (215 K) [Fig.2(e)]. 

To map the in-plane magnetic anisotropy and the easy axis of the heterostructure, the in-plane angle $\phi$ (angle between in-plane applied magnetic field and sample [100] direction) dependent MOKE measurements at room temperature (T-phase of the BTO substrate) were performed in the 0$^\circ$-360$^\circ$ angular range as shown by the schematic in Fig.2(f). At that temperature, we observe that the easy axis of the system lies along the [110] and [1-10] directions with four-fold anisotropy as shown in Fig.2(g).

 These distinct changes in the M(H) behavior, especially the shape of M(H) curves (the demagnetization process), prove that strain induced-lattice distortions across the successive structural phase transitions crucially affect the magnetic anisotropy and modify the APBs density in the magnetoelastically coupled Fe$_3$O$_4$/BTO heterostructure. 

\subsection{Electrical transport properties}

\begin{figure*}[t]
\includegraphics[width=.89\textwidth]{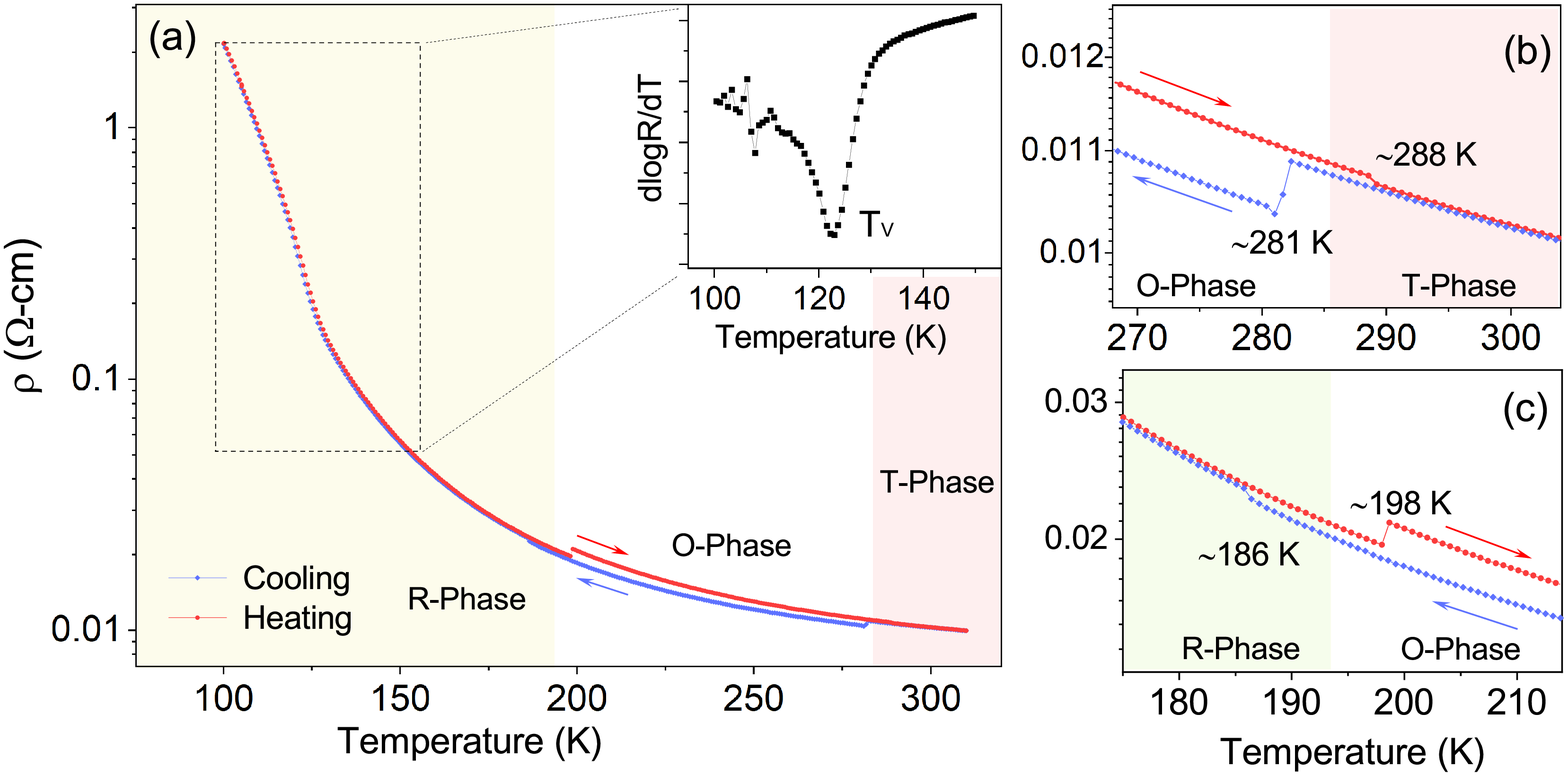}
\caption{(a) Resistivity $\rho$ of Fe$_3$O$_4$/BTO(001) heterostructure as function of temperature. The inset shows dlogR/dT. (b,c) show the close view of the resistivity jumps across the T-O and O-R structural phase transitions of the BTO substrate, respectively.}
\label{fig3}
\end{figure*}

Figure 3 shows the temperature dependent resistivity $\rho$(T) behavior between 100-310 K of the Fe$_3$O$_4$/BTO(001) heterostructure in heating (red) and cooling (blue) cycles. This resistivity behavior exhibits a metal-to-insulator transition below 123 K with significantly higher resistivity values, indicating the characteristic Verwey transition at 123 K, which is clearly seen in the dlogR/dT plot in the inset of Fig.3(a). This confirms the formation of a highly stoichiometric Fe$_3$O$_4$ thin film. Figure 3(a)clearly shows the discontinuous drop (at 281 K) and jump (at 186 K) in the cooling cycle of $\rho$(T), induced via strain at the BTO substrate tetragonal to orthorhombic (T-O) and orthorhombic to rhombohedral (O-R) structural phases transitions, respectively. In the warming cycle, these discontinuities (at 288 K and 198 K) change sign and appear at somewhat higher temperatures because of thermal hysteresis associated with the first order structural phase transitions. These discontinuities in $\rho$(T) are fully consistent with our magnetization $M(T)$ results and further highlight that electrical transport and magnetic properties of the \fo\ thin film are very sensitive to the induced strain across the structural phase transition of BTO.

\subsection{Electrical control of magnetic properties}

\begin{figure*}[t]
\centering
\includegraphics[width=1\textwidth,keepaspectratio]{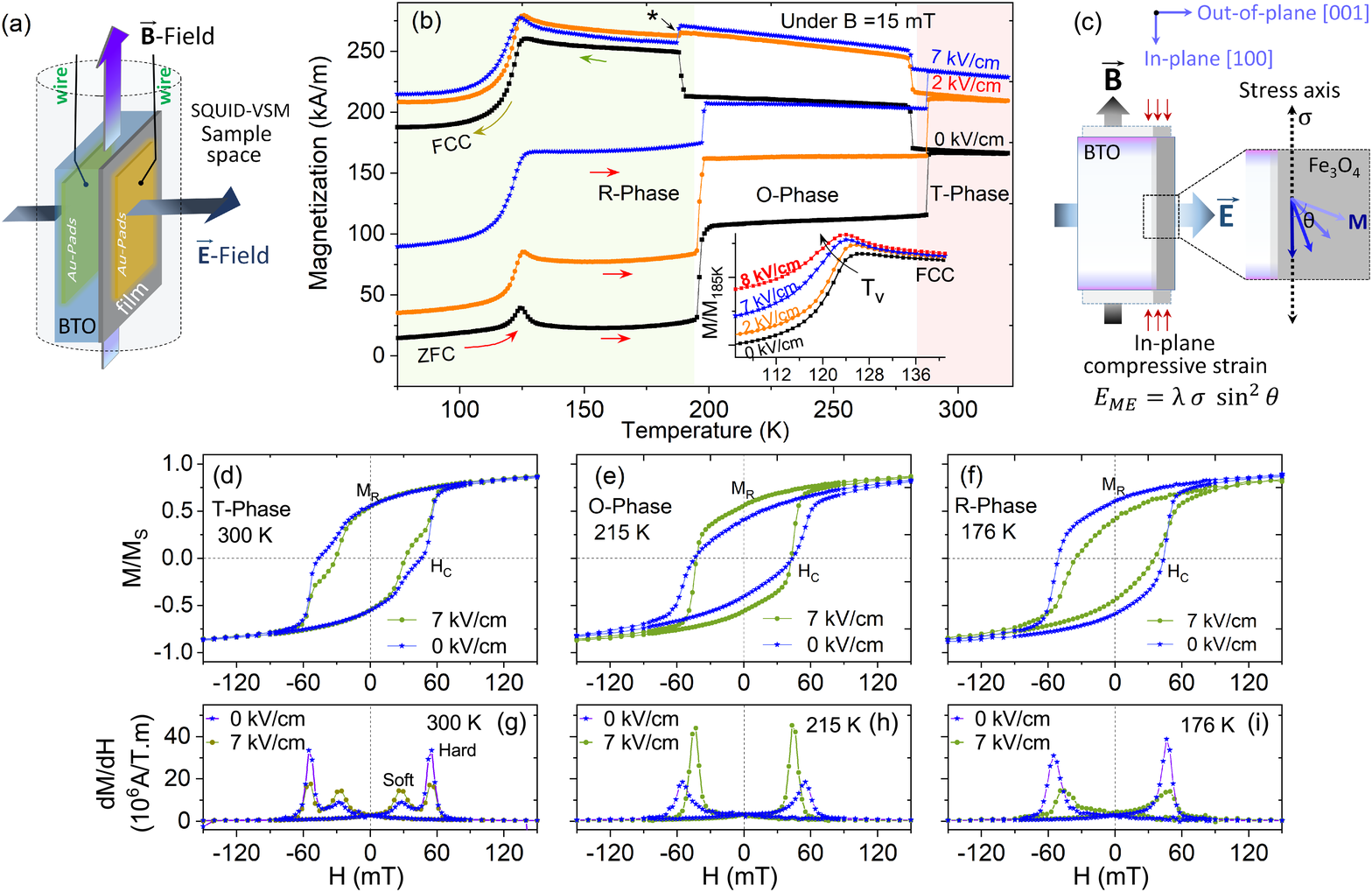}
\caption{
(a) Schematic representation of magnetic and electric field directions during the $in$-$situ$ electric-field-dependent magnetization measurements. FCC-ZFC M(T) behavior of Fe$_3$O$_4$ thin film under (b) $B$ = 15 mT applied magnetic field (in-plane) and with $E$ = 0, 2 and 7 kV/cm applied electric field (out-of-plane), the inset shows a close view of the normalized $M/M_{\text{185 K}}$ versus temperature behavior around the Verwey transition. (c) Schematic representation of switching of magnetization direction via electric field induced stress in the Fe$_3$O$_4$ thin film. Normalized M(H) hysteresis/dM/dH of Fe$_3$O$_4$ thin film recorded under 0 and 7 kV/cm applied electric field at (d)/(g) 300 K, (e)/(h) 215 K, (f)/(i) 176 K corresponding to different structural phases of underneath BTO (001) substrate.}
\label{fig4}
\end{figure*}

So far it is apparent that the thermally driven structural phase transitions strongly influence the magnetic properties of the Fe$_3$O$_4$ layer in the magnetoelastically coupled Fe$_3$O$_4$/BTO heterostructure. Now we focus on the control of the magnetic properties such as the magnetic anisotropy and Verwey transition of the Fe$_3$O$_4$ thin film by the application of an $in$-$situ$ static electric field ($E$-field). We thus have performed M(T) measurements under applied E-fields up to 8 kV/cm. The E-field was applied across the 0.5 mm thick BTO substrate along its [001] direction by depositing thin gold pads on both sides of the Fe$_3$O$_4$/BTO sample as shown in the schematic in Fig.4(a). The M(T) behavior under an applied magnetic field of B = 15 mT in-plane ($ab$-plane of film) at 0, 2 and 7 kV/cm applied E-fields (perpendicular to both magnetic field and to the plane of film) were again collected under FCC and ZFC conditions, and are shown in Fig.4(b). 

In the FCC curves, it can be clearly seen that the magnetization jumps across the structural transitions observed in zero field are drastically modified under the application of an E-field. For $E$=2 kV/cm, the room temperature (300 K) magnetization is enhanced by 25 \% with respect to $E$=0 and the height of the magnetization jump is reduced across the T-O-phase transition, while being almost completely suppressed and changed in sign at the O-R transition. For a further increase to $E$=7 kV/cm, the magnetization jump height across the T-O phase transition further reduces. However, more interestingly, it increases across the O-R phase transition [indicated as * in Fig.4(b)] and the sign change compared to zero applied E-field becomes clearly visible. The magnitude of these magnetization jumps in the ZFC cycles between T-O and O-R phase transitions monotonically decreases with increasing E-field. These observations clearly point towards an E-field driven magnetic anisotropy.

To better understand these discontinuities and  the E-field tunable magnetic anisotropy, we have recorded the E-field dependent M(H) hystereses in $E$=0 kV/cm and $E$=7 kV/cm, which are shown in Figs.4(d-f). In comparison to the earlier discussed zero E-field hysteresis (section III-B), the M(H) behavior at $E$ = 7 kV/cm exhibits a modified double coercivity, which starts to disappear and the overall coercivity of the heterostructure appears to be reduced with unaffected remanence ratio M$_R$/M$_S$ as shown in Fig.4(d). This reduction of coercivity suggests the reduction of hard phase, as the application of an E-field in the tetragonal phase leads to a transformation from a multidomain  (ferroelectric) state (assosiated with hard and soft magnetic phases of Fe$_3$O$_4$) to a single domain state, which favors fewer grain boundaries and pinning sites. This is clearly visible in dM/dH shown in Fig.4(g), which show almost identical coercive fields but different intensities.

In the orthorhombic (O) phase at 215 K, the application of a $E$ = 7 kV/cm field leads to a 37\% higher degree of squareness (M$_R$/M$_S$) and modify the corecivity (dM/dH) of M(H) [see Fig.4(e) and 4(h)], as consequence of the strain induced variation of the magnetic anisotropy. By the application of such a high E-field (typically, a 1 kV/cm E-field is high enough to align most of the domains along the E-field direction\cite{Tazaki2009}), the unit cell of BTO gets elongated in the direction of the applied E-field and produces an in-plane compressive strain (positive longitudinal \textit{d}$_{33}$ and negative transverse \textit{d}$_{31}$ piezoelectric coefficients\cite{Tazaki2009}) as shown by the schematic in Fig.4(c). This piezo-strain leads to a magnetic anisotropy modification in \fo/BTO through magnetoelastic coupling.

In magnetoelastically coupled heterostructures, the magnetoelastic energy ($E_{\text{ME}}$) that gives rise to a uniaxial anisotropy is given by the expression: 

\[E_{\text{ME}} = (3/2) \lambda \sigma \sin^{2}(\theta),\]

where $\lambda$, $\sigma$ and $\theta$ are the magnetostriction constant, stress, and angle between the spontaneous magnetization and  the stress axis\cite{ME}, as shown schematically in  Fig.4(c). For \fo, the magnetostriction constant is negative\cite{tian2008} ($\lambda<$ 0) which implies that an $E$-field induced compressive strain ($\sigma<$ 0) favors $\theta$ = 0, which corresponds to an in-plane alignment of the moments. In other words, the E-field induced compressive strain within the orthorhombic phase forces the easy axis (see Fig.4(g)) into the film's $ab$-plane. This is in agreement with the strongly enhanced M$_R$/M$_S$ ratio in M(H) at 215 K we observe in Fig.4(e). In contrast, in the R-phase of BTO at 176 K, the degree of squareness M$_R$/M$_S$ and coercivity at $E$ = 7 kV/cm is reduced (Fig.4(f) and 4(i)) with respect to $E=0$ and displays a reduced coercivity. This in turns implies that the easy axis (see Fig.4(g)) now moves away from the plane of  the sample and the system exhibits an E-field induced perpendicular magnetic anisotropy in this R-phase. This E-field induced reorientation behavior is now opposite to what is observed in the O-phase, which is clearly consistent with  the observed sign reversal of the magnetization jump at high E-fields at  the O-R phase transition in our M(T) data.

The magnetization data shown in Fig. 4(b) reveal a further interesting feature as function of applied electric field. The Verwey transition temperature $T_V$ of \fo, which is characteristic of the charge ordered state and very sensitive to the stoichiometry of \fo, ferroelastic strain and densities of APBs\cite{Markovich2002}, is strongly affected by the application of the electric field. Under the application of $E$-fields up to 8 kV/cm, it significantly shifts (by 4 K from 123 K) towards lower temperature as displayed in the normalized FCC magnetization M/M$_{185 K}$ in the inset of Fig.4(b). The profile of the transition is drastically modified in the ZFC cycle. The bifurcation between the FCC-ZFC cycles is strongly reduced below $T_V$ under the application of an $E$-field [Fig.4(b)], whereas it was increasing under zero electric field with increased magnetic field from 15 mT to 30 mT as discussed above in Fig. 2(a). 

Recently, Kumar \textit{et al.}\cite{AKumar2018} reported that during the deposition of a \fo/MgO(001) heterostructure, the electric field plays a crucial role in the control of the density of APBs and the antiferromagnetic (AFM) interaction across the APBs in the $E$-field assisted growth of their \fo\ thin films. Therefore, under the application of an $E$-field and different strain states, the density of APBs can be modified during the temperature cycling in the recording of  the M(T) behavior and can be considered as microscopic origin of the systematic variation in the profile of Verwey transition that we observe. However, in our case of the \fo/BTO(001) heterostructure, it is difficult to disentangle the individual contributions of the strain and the density of the APBs.

\subsection{Electric field dependent Raman study}

\begin{figure*}[t]
\includegraphics[width=0.65\textwidth]{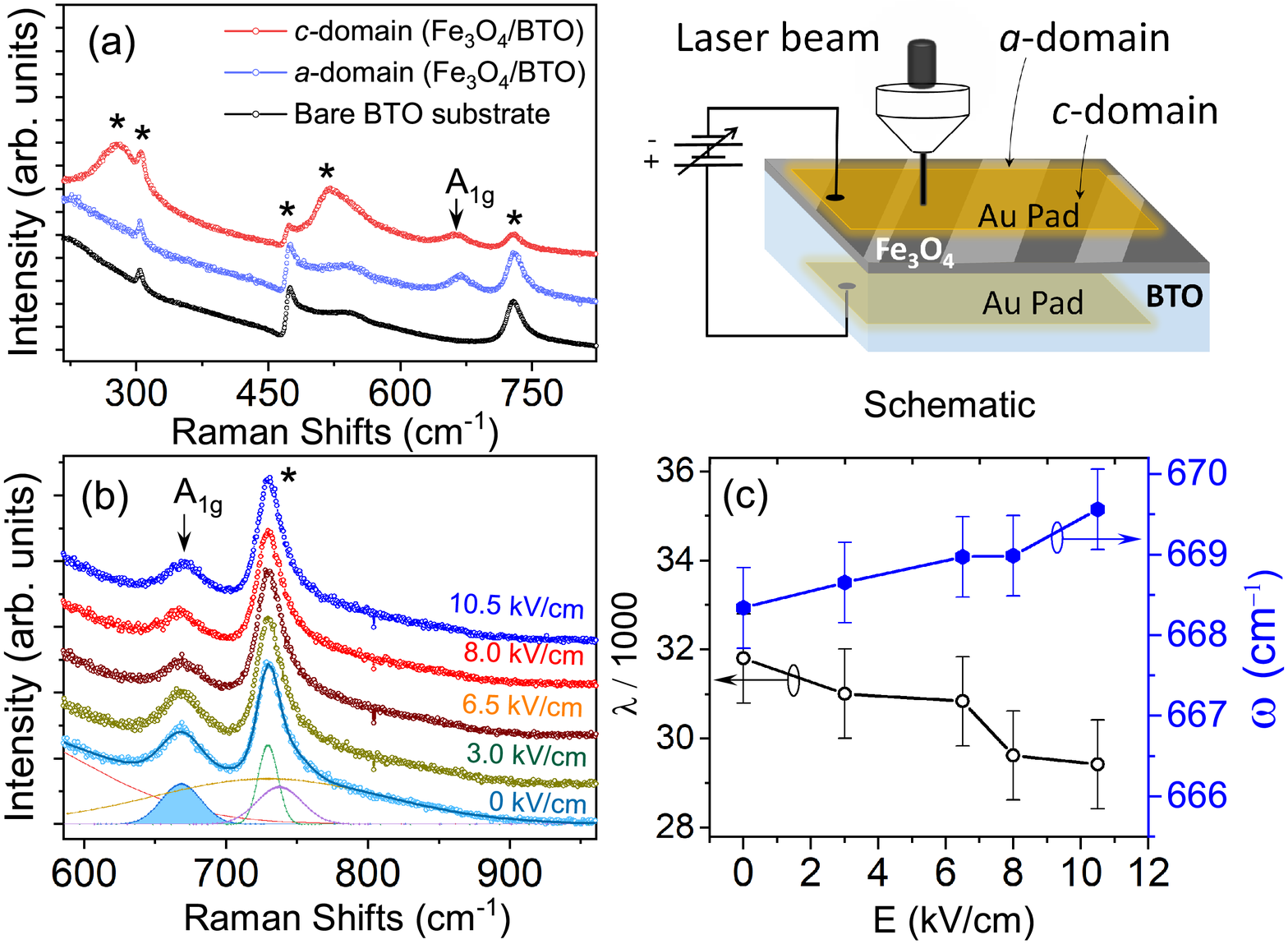}
\caption{Schematic of $E$-field dependent Raman measurements. (a) Room temperature Raman spectra of  the Fe$_3$O$_4$/BTO heterostructure corresponding to \textit{a} and $c$-domains along with the reference bare BTO substrate. The asterisks indicate the Raman modes of BTO, the arrow indicates the most intense A$_{1g}$ Raman mode of Fe$_3$O$_4$. (b) Fitting of A$_{1g}$ mode by Gaussian lineshapes. (c) Variation of spin-phonon coupling constant $\lambda$ and frequency $\omega$ of A$_{1g}$ Raman mode as function of E-field.}
\label{fig5}
\end{figure*}

To gain insights into the influence of the electric field on the presence of APBs and on the lattice distortions, room temperature unpolarized micro-Raman measurements were performed for different $E$-fields in the frequency range of 100 to 1000 cm$^{-1}$ as shown in the schematic of Fig.5. Raman spectra of our Fe$_3$O$_4$/BTO heterostructures, corresponding to $a$ and $c$-domains along with data of a reference bare BTO substrate for comparison are shown in Fig. 5(a). Raman modes corresponding to  the underneath BTO substrate are indicated by asterisks.
In the Fe$_3$O$_4$/BTO magnetic heterostructure, only the most intense Raman active mode A$_{1g}$ at 668 cm$^{-1}$ (indicated by arrow) is observed. The T$_{2g}$ mode of Fe$_3$O$_4$ is not seen, because of its very weak intensity and overlap with the [B$_1$+ E(TO+LO)] modes (306 cm$^{-1}$) of  the BTO substrate.\cite{Shiratori2007} In thin film samples of \fo, these two modes, A$_{1g}$ and T$_{2g}$, are very sensitive to the strain state and the electronic properties of Fe$_3$O$_4$. The observed A$_{1g}$ mode has a slightly different line width and mode position on  the two different domains, which again reveals that the Fe$_3$O$_4$ thin film is elastically coupled to the 90$^\circ$ $a$-$c$ domain pattern of the underlying BTO substrate.
The observed Raman spectra are fitted by a sum of Gaussian lineshape functions [Fig.5(b)], in order to evaluate the spin-phonon coupling constant ($\lambda$) for the A$_{1g}$ Raman mode given by Allen’s formula:\cite{Allen}

\begin{center}
\[\lambda_{i}= \frac{{\Gamma_{i}}}{\omega^2} \frac{{g_{i}}}{2\pi N(E)},
\]
\end{center}
where $\omega$ is  the angular frequency, $\Gamma$ is  the linewidth (full width at half-maximum extracted from Gaussian fits), $N(E)$ is the density of states at the Fermi level (3 states/eV per Fe atom for \fo\ at room temperature)\cite{Zhang91} and $g_i$ is the degeneracy of the $i^{\text{th}}$ mode.

The values of $\lambda$ are directly related to the density of APBs, and found to be smaller for  the A$_{1g}$ mode than for the T$_{2g}$ mode in previously reported \cite{Tiwari2008,Phase2006} heterostructures when compared to \fo thin films
grown on Si substrates or bulk \fo, which is nearly devoid of APBs.

In our sample, the value of $\lambda$ for the A$_{1g}$ mode without applying any E-field is 0.0318, which is much smaller than the \fo films grown on Si substrates and indicatetes that the grown film are rather free from APBs.\cite{Tiwari2008} However, it is difficult to make a conclusive statement about the APB density based on the value of $\lambda$ for the A$_{1g}$ mode alone, because both $\lambda$s for A$_{1g}$ and for T$_{2g}$ should be considered. In fact, the $\lambda$ value for the T$_{2g}$ mode is expected to being strongly affected by the variation of APBs and would thus provide a better picture of the APB density in the \fo\ thin film.\cite{Tiwari2008} Nevertheless, our observed value of $\lambda$ for  the A$_{1g}$ mode is fully consistent with the formerly reported values\cite{AKumar2018, Tiwari2008} and clearly points out that the number of APBs/grain boundary defects is rather small and can be induced by 90$^\circ$ ferroelectric $a$-$c$ stripe domains and elastic strain. Our findings are also consistent with an earlier study by Kumar \textit{et al.}\cite{AKumar2018}, who reported that  the E-field during the growth of \fo\ thin films plays a critical role in  the control of the density of APBs. For our Fe$_3$O$_4$/BTO heterostructure, the variation in the value of $\lambda$ for the A$_{1g}$ Raman mode as function of different applied E-fields is shown in Fig.5(c). We observe that the value of $\lambda$ is only slightly decreasing with applied E-field and change by only about 7 \% up to the highest applied fields of $E$= 10.5 kV/cm. These minor changes indicate that the density of APBs in the epitaxial thin film cannot be easily modified once the thin film is thermodynamically stable after the deposition. Therefore, the observed modifications in the magnetic properties are then mainly caused by strain induced lattice distortions and  the magnetic anisotropy.

\section{Conclusions}
In conclusion, an epitaxial \fo/BaTiO$_3$ (001) heterostructure was prepared by pulsed laser deposition. RSM measurements show that the BTO $a$-$c$ stripes domain are imprinted in the Fe$_3$O$_4$ thin film layer in the form of bilateral domains. Sharp discontinuous jumps in resistivity and magnetization along with large variations in coercivity are observed across the thermally driven characteristic structural phase transitions of BTO and reveal the presence of strong magnetoelastic coupling in the Fe$_3$O$_4$/BTO heterostructure. \textit{In-situ} electric field dependent magnetization measurements successfully demonstrate the electrical control of the magnetization easy axis and of the Verwey metal-insulator transition of \fo\ in a wide range of temperatures between 20-300 K. Room temperature micro-Raman spectra under different applied electric fields show that the modification in the magnetization and Verwey transition are mainly caused by strain induced lattice distortions and the magnetic anisotropy, rather than by an alteration in APBs defect density. These results provide a pathway for tailoring the structural, magnetic and electronic transport properties in such magnetic thin film systems through controlled strain transmission either induced by an electric field across the FE-BTO interface via magnetoelectric coupling or by magnetoelastic coupling using the structural phase transitions of BTO, which has potential impacts on the design of future spintronic devices.

\acknowledgments
G.P is thankful to Dr. V. R. Reddy, Er. Anil Gome for RSM measurements and Dr. Zaineb Hussain for fruitful discussion.

\bigskip
%Create the reference section using BibTeX:
\bibliography{Bibfile1}

\end{document}